\documentclass{osa-article}

\journal{osajournal}

\articletype{Research Article}

\usepackage[version=4]{mhchem}
\usepackage{amsmath,commath,siunitx,braket,physics,float,multirow,booktabs}
\usepackage{titling}

\newcommand{\SIadj}[2]{\SI[number-unit-product={\text{-}}]{#1}{#2}}
\newcommand{\SIrangeadj}[3]{\numrange[range-phrase = --, range-units = single]{#1}{#2}-\si{#3}}

\newcommand*{\E}{\mathrm{e}}
\newcommand*\diff{\mathop{}\!\mathrm{d}}

\newcommand{\etal}{\textit{et\,al}.\@ }
\newcommand{\eg}{\textit{e}.\textit{g}., }

\newcommand*{\rttensor}[1]{\overline{\overline{#1}}}

\begin{document}

\title{Efficient soliton self-frequency shift in hydrogen-filled hollow-core fiber}

\author{Yi-Hao Chen,\authormark{1,*} and Pavel Sidorenko\authormark{1} and Enrique Antonio-Lopez\authormark{2} and Rodrigo Amezcua-Correa\authormark{2} and Frank Wise\authormark{1}}

\address{\authormark{1}School of Applied and Engineering Physics, Cornell University, Ithaca NY 14853, USA\\
\authormark{2}University of Central Florida, CREOL, The College of Optics and Photonics, Orlando, Florida 32816, USA}

\email{\authormark{*}yc2368@cornell.edu} 



\begin{abstract}
We report a study of soliton self-frequency shifting in hydrogen-filled hollow-core fiber. The combination of hydrogen and short \SIadj{40}{\fs} input pulses underlies clean and efficient generation of Raman solitons between \num{1080} and \SI{1600}{\nm}. With \SIadj{240}{\nano\joule} input pulses, the Raman soliton energy ranges from \num{110} to \SI{20}{\nano\joule} over that wavelength range, and the pulse duration is approximately \SI{45}{\fs}. In particular, \SIadj{70}{\nano\joule} and \SIadj{42}{\fs} pulses are generated at \SI{1300}{\nm}. Numerical simulations agree reasonably well with experiments and predict that microjoule-energy tunable pulses should be possible with higher-energy input pulses.
\end{abstract}

Recently, hollow-core fibers have attracted a lot of attention due to the abundant physics that is possible when pulses inside a fiber interact with various gases. With inert gases, only the electronic Kerr response takes place, and such fibers have been used in applications such as ultraviolet generation \cite{Joly2011}, photoionization-induced blue-shift \cite{Saleh2011}, pulse compression, and supercontinuum generation \cite{Adamu2019,Travers2019,Elu2021}. In molecular gases, Raman effects are important. In contrast to solid-core silica fibers, Raman scattering in molecular gases has a long dephasing time (\SI{\sim100}{\ps} depending on the gas pressure) \cite{Weber1994}; this underlies interesting phenomena that are unique to gases. Raman-enhanced Kerr nonlinearity \cite{Belli2015} and solid-state physics in a gas-induced temporal crystal \cite{Saleh2015}, just to name a few effects, have been demonstrated. Loranger \etal showed that \SIadj{40}{\fs} pulses at \SI{1800}{\nm} can be generated by launching \SI{300}{\fs} pulses into a hydrogen-filled hollow-core fiber \cite{Loranger2020a}. Initial observations of the soliton self-frequency shift (SSFS) in gas-filled hollow-core fibers reported small spectral shifts of tens of nanometers \cite{Ouzounov2003,Luan2004,Gerome2008}. Stimulated Raman scattering plays a role in several recent studies, especially those of continuum generation and soliton self-compression \cite{Carpeggiani2020,Beetar2020}. The interplay of Raman and multimode effects has led to the recent observation of multidimensional solitary states \cite{Safaei2020}. Despite the importance of Raman effects, there is no systematic study of SSFS in hollow-core fibers to the best of our knowledge.

SSFS has long been a good candidate for wavelength-tunable fiber sources \cite{Dekker2011,Bi2016,Wang2011}. In particular, nonlinear microscopy, such as three-photon imaging, requires high peak power at \num{1300} and \SI{1700}{\nm} to overcome the depth limit of two-photon imaging \cite{Ouzounov2017,Chow2020}. SSFS occurs when intrapulse Raman scattering continuously transfers energy from the high-frequency part of a soliton to the low-frequency part in an anomalous-dispersion medium \cite{Gordon1986,Mitschke1986}. The Raman soliton thus gradually shifts to the red as it propagates. Highly nonlinear photonic crystal fibers \cite{Chan2008} and fibers made with different materials, \eg tellurite glasses \cite{Bi2016}, were used to produce Raman solitons in various wavelength ranges but with limited energies. The use of large-mode-area (LMA) silica fiber allowed the generation of pulses tunable from \num{1580} to \SI{2130}{\nm}, with up to \SIadj{45}{\nano\joule} pulse energy and \SIadj{70}{\fs} pulse duration \cite{Wang2011}. With a solid-core anti-resonant photonic crystal fiber, \SIadj{95}{\nano\joule} and \SIadj{85}{\fs} pulses at \SI{1800}{\nm} were generated \cite{Delahaye2021}. The process of soliton self-mode conversion (SSMC) exploits higher-order modes of multimode fiber to achieve anomalous dispersion and large mode area at wavelengths below \SI{1300}{\nm} \cite{Rishoej2019}. SSMC has yielded \SIadj{80}{\nano\joule} and \SIadj{74}{\fs} pulses at \SI{1300}{\nm}. Although they have impressive peak power, further scaling of the pulse energy in solid-glass fibers appears to be difficult.

In contrast to solid-core fibers, nonlinearity of gases is low and high-energy solitons can propagate through a gas-filled fiber with a low soliton number. In addition, the pressure-tunable dispersion and broad transmission bands \cite{Adamu2019} of anti-resonant hollow-core fibers (AR-HCFs) make this platform extremely attractive for high-energy SSFS.

In this letter, we demonstrate SSFS in a hydrogen-filled AR-HCF. Thanks to the use of hydrogen and short input pulses, efficient and clean SSFS occurs. Continuous tuning of the wavelength between \num{1080} and \SI{1600}{\nm} is observed. Pulse energies in the range of \num{20} to \SI{110}{\nano\joule} and durations below \SI{50}{\fs} are obtained over this spectral range. Numerical simulations account well for the experimental results and predict scaling of the process to microjoule energies with stronger pump pulses.

To generate a clean Raman soliton with high efficiency, both the soliton number of the launched pulse and the number of participating Raman transitions should be as small as possible. Several studies have obtained wavelength-tunable sources by extracting the reddest lobe from a supercontinuum \cite{Carpeggiani2020,Fan2020,Beetar2020}; however, they failed to meet the above conditions and had relatively low efficiency. For example, with \ce{N2}, Carpeggiani \etal demonstrated \SI{57}{\fs} pulses with \SI{8}{\percent} efficiency in the \SIrangeadj{1450}{1650}{\nm} spectral window. The strong launched pulse evolves as a high-order soliton with a soliton number $N$ given by
\begin{equation}
N^2=\frac{\gamma P_0T_0^2}{\abs{\beta_2}}\approx\frac{\gamma E_0T_0}{\abs{\beta_2}}
\label{eq:N}
\end{equation}
where $\gamma$ is the nonlinear coefficient, $P_0$ is the peak power, $T_0$ is the pulse duration, $E_0\approx P_0T_0$ is the pulse energy, and $\beta_2$ is the dispersion. Perturbations from higher-order dispersion, self-steepening, etc., cause the field to undergo fission into $N$ constituent fundamental solitons. A pulse of a smaller soliton number breaks into fewer fundamental solitons, so the reddest Raman soliton takes up more energy. Soliton number is directly proportional to the pulse duration, so a shorter input pulse will lead to a more efficient SSFS process. In addition, pulse bandwidth needs to be large enough to support the SSFS. Recently, a new fiber amplification regime that allows the generation of pulses with spectra well beyond the gain-narrowing limit was demonstrated \cite{Sidorenko2019,Sidorenko2020}. With LMA \ce{Yb}-doped fiber, such a gain-managed amplifier can deliver \SIadj{1}{\micro\joule} and \SI{40}{\fs} pulses. The short pulse duration makes it an ideal simple source for investigation of SSFS with small soliton number. We choose \ce{H2} over other gases to minimize the number of Raman transitions that can occur. Because it has a large energy difference between energy states, most of the population stays in the lowest possible states at room temperature. Only two rotational transitions, S(0) with para-\ce{H2} ($\triangle\nu=0, J=0\rightarrow2$) and S(1) with ortho-\ce{H2} ($\triangle\nu=0, J=1\rightarrow3$), and one vibrational Raman transition, Q(0) ($\nu=0\rightarrow1, \triangle J=0$), play important roles. Because the ratio of ortho-\ce{H2} to para-\ce{H2} populations is 3:1 at room temperature, S(1) dominates over S(0). Despite the larger Raman gain of the vibrational Raman transitions than the rotational ones, self-seeded SSFS of a sub-\SIadj{50}{\fs} broadband pulse dominates over the noise-seeded discrete vibrational Raman transitions. As mentioned above, several studies have investigated Raman-shifting in nitrogen-filled fibers. With \ce{N2}, multiple rotational Raman transitions come into play, which increases the pulse duration and reduces the efficiency of conversion to a single wavelength (see the discussion of nitrogen-filled AR-HCF in Supplement 1).

The experimental setup is depicted in Fig.~\ref{fig:schematic}(a). A gain-managed amplifier supplies \SIadj{400}{\nano\joule} and \SIadj{32}{\fs} pulses [Fig.~\ref{fig:schematic}(b) and \ref{fig:schematic}(c)]. The linearly-polarized pulses are coupled, with \SI{60}{\percent} efficiency, into a \SIadj{2}{\m} AR-HCF with a \SIadj{30}{\micro\m} core diameter and \SIadj{300}{\nm} wall thickness of its tubes [Fig.~\ref{fig:schematic}(d)]. By paying more attention to the coupling optics, coupling up to \SI{90}{\percent} should be achievable. The fiber is designed to have less than \SIadj{0.3}{\dB/\m} loss between \num{800} and \SI{1700}{\nm} [Fig.~\ref{fig:schematic}(e)] \cite{Habib2019}. It maintains anomalous dispersion at wavelengths above \SI{1}{\micro\m} for hydrogen pressures up to \SI{100}{\bar} [Fig~\ref{fig:schematic}(f)] \cite{Bache2019}. Propagation in the fundamental transverse mode of the fiber was confirmed (see Fig.~S2 in Supplement 1).

\begin{figure}[h!]
\centering
\includegraphics[width=0.82\linewidth]{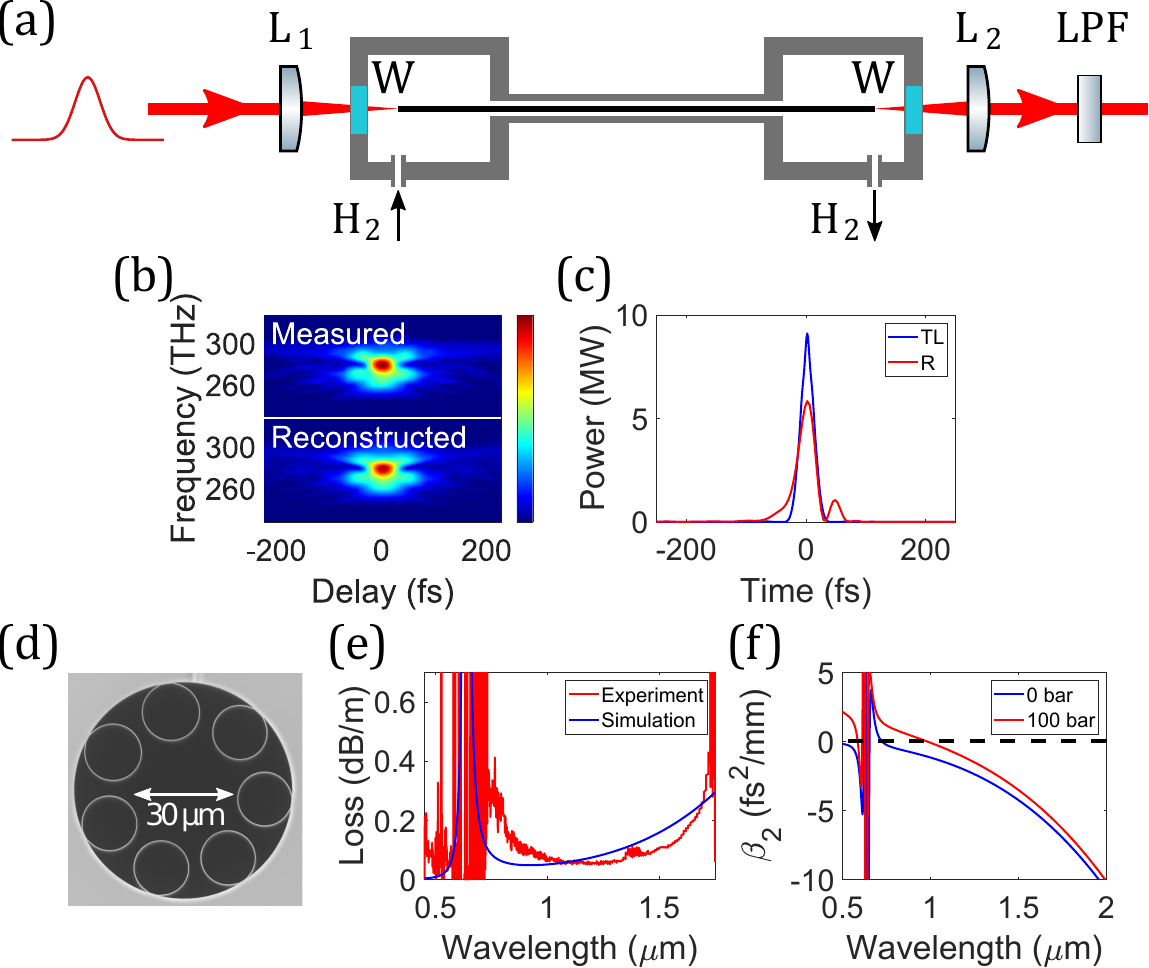}
\caption{(a) Schematic of the gas system. Two gas cells are connected with a stainless-steel tube where the AR-HCF lies. $L_1$ and $L_2$ are focusing and collimating lenses with \SIadj{50}{\mm} focal length; {LPF} is a long-pass filter for selecting the Raman soliton; $W$ is a sapphire window. (b) Measured and reconstructed FROG traces of the input pulse. (c) Temporal profiles of the retrieved (R) and transform-limited (TL) input pulses. Measured (d) cross section and (e) loss curve of the seven-tube AR-HCF. (f) Calculated dispersion curves of the fiber with vacuum and \SIadj{100}{\bar} hydrogen pressure. The zero-dispersion wavelengths are \SI{720}{\nm} and \SI{980}{\nm}, respectively.}
\label{fig:schematic}
\end{figure}

The experimental results are summarized in Fig.~\ref{fig:exp_spectrum}. The measured output spectra [Fig.~\ref{fig:exp_spectrum}(a)] exhibit a red-shifting peak on the long-wavelength side of the spectra, which contributes entirely to a single Raman soliton, as expected owing to the excitation of primarily a single rotational Raman transition. The frequency shift increases roughly linearly with \ce{H2} pressure up to about \SI{90}{\bar}, where it seems to saturate. At \SI{98.2}{\bar}, we observe a soliton at \SI{1600}{\nm}. After isolating the Raman soliton with a long-pass spectral filter, we measure its pulse duration. The pulse duration measured by frequency-resolved optical gating (FROG) is around \SI{45}{\fs} for all pressures in the range investigated [Fig.~\ref{fig:exp_spectrum}(b)]. The pulse energy varies from above \SI{100}{\nano\joule} for small frequency shifts to \SI{20}{\nano\joule} at the largest shift. The peak power of the Raman soliton thus exceeds \SI{1}{\MW} for pressures (wavelengths) up to \SIadj{50}{\bar} (\SI{1430}{\nm}). With \SIadj{40}{\bar} \ce{H2} pressure, we obtain \SIadj{70}{\nano\joule} and \SIadj{42}{\fs} pulses at \SI{1300}{\nm}, for a peak power of nearly \SI{2}{\MW} at this biologically-important wavelength.

In addition to the Raman soliton, we observe dispersive waves in the visible spectral region for gas pressures between \num{50} and \SI{80}{\bar}. The sharp dispersion slope near the resonance enables phase-matched dispersive-wave generation despite the high confinement loss \cite{Tani2018} (see the discussion of dispersive waves in Supplement 1).

\begin{figure}[h!]
\centering
\includegraphics[width=0.95\linewidth]{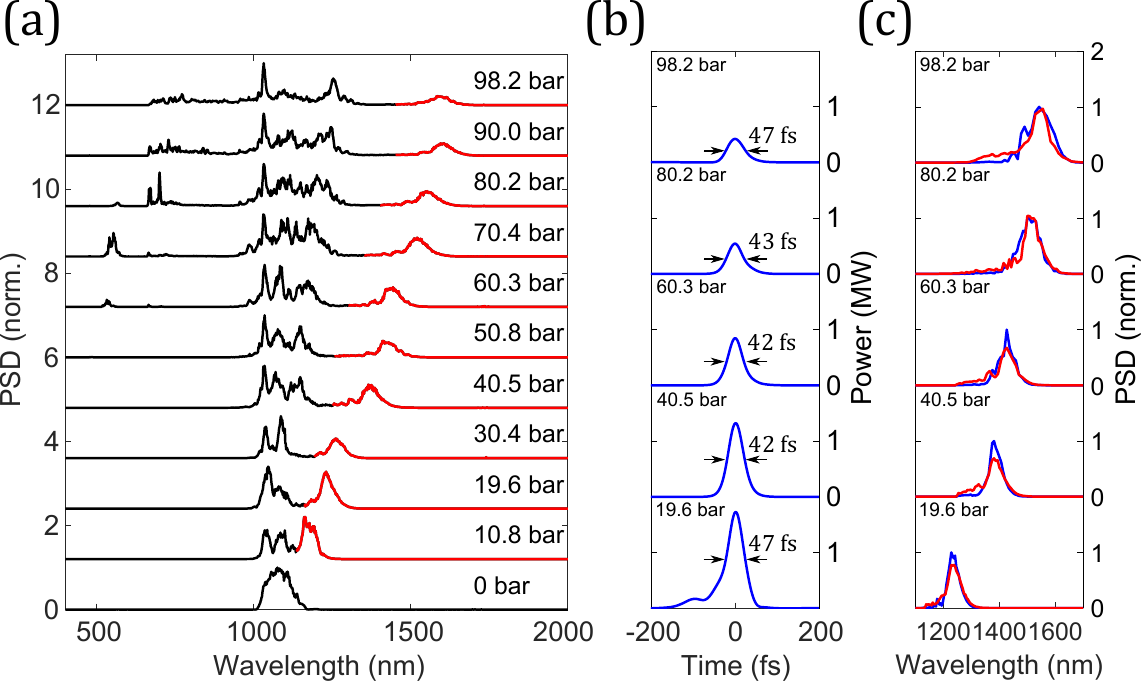}
\caption{(a) Output spectra (red curve corresponds to the band transmitted by the long-pass filer). {PSD}: power spectral density. (b) Raman pulses measured with FROG at indicated pressures. (c) Spectra retrieved with FROG (blue) and measured with an optical spectrum analyzer (red).}
\label{fig:exp_spectrum}
\end{figure}

To obtain insight into the experimental results, we numerically simulated the process with a unidirectional pulse propagation equation that includes the electronic and Raman nonlinearities of \ce{H2} (details are in Supplement 1),
\begin{align}
& \partial_zA(z,\Omega)=i\left[\beta(\omega)-\left(\beta_0+\beta_1\Omega\right)\right]A(z,\Omega)+ \nonumber \\
& \frac{i\omega}{4}Q^R_{1111}\left(\frac{3\epsilon_0\chi^{(3)}_{\text{electronic}}}{4}\mathfrak{F}[\abs{A}^2A]+\mathfrak{F}\left[A\left[R(t)\ast \left(\abs{A}^2\right)\right]\right]\right),
\end{align}
where $z$ is the propagation distance, $\omega$ and $\Omega=\omega-\omega_0$ are the angular frequency and the frequency offset from the center of the frequency window, $A(z,\Omega)$ is the fundamental-mode electric field, $\beta(\omega)$ is the propagation constant including the effect of the gas, $\beta_0$ and $\beta_1$ are the free parameters of the model where $\beta_1=1/v_g$ represents the inverse moving speed of the reference frame in simulations. We apply the simplified model proposed by Bache \etal to compute both the loss and the propagation constant of the propagating mode \cite{Bache2019}. $Q^R_{1111}=\frac{4}{\epsilon_0^2n_{\text{eff}}^2c^2}\frac{1}{A_{\text{eff}}}$ in which $n_{\text{eff}}=\frac{\beta(\omega)}{k_0}$ and $A_{\text{eff}}$ represent the effective refractive index and the effective area of the mode. $\chi^{(3)}_{\text{electronic}}$ is the Kerr nonlinear coefficient. The total Raman response includes vibrational and rotational parts, $R(t)=R^{\text{rot}}(t)+R^{\text{vib}}(t)$. The vibrational response
\begin{align}
& R^{\text{vib}}=N_g\frac{1}{4\mu}\E^{-\gamma_2^{\text{vib}}t}\times \nonumber \\
& \sum_J(2J+1)\rho^{(0)}_J\frac{\left(\od{\alpha}{\mathbb{Q}}\right)^2_0+\frac{4}{45}\frac{J(J+1)}{(2J-1)(2J+3)}\left(\od{\left(\triangle\alpha\right)}{\mathbb{Q}}\right)^2_0}{\omega_{1J,0J}}\sin\left(\omega_{1J,0J}t\right)
\end{align}
and the rotational response
\begin{align}
R^{\text{rot}} & =N_g\frac{1}{15\hbar}\left(\triangle\alpha\right)^2\E^{-\gamma_2^{\text{rot}}t}\times \nonumber \\
&\sum_J\left(\rho^{(0)}_J-\rho^{(0)}_{J+2}\right)\frac{(J+2)(J+1)}{2J+3}\sin\left(\omega_{0J+2,0J}t\right),
\end{align}
are both summed over the rotational quantum number $J$. $N_g$ is the number density, $\mu$ is the reduced mass of the gas molecule, $\gamma_2$ is the dephasing rate, $\rho^{(0)}_J$ is the Boltzmann-distributed population of the energy state ($\nu=0, J$) without the perturbed electric field, $\left(\frac{\diff\alpha}{\diff\mathbb{Q}}\right)_0$ is the derivative of the gas mean polarizability with respect to the normal coordinate of the molecule at equilibrium, $\triangle\alpha$ is the polarizability anisotropy, and $\omega_{\nu_1J_1,\nu_2J_2}=\triangle E_{\nu_1J_1,\nu_2J_2}/\hbar$ represents the energy difference between two states.

Figs.~\ref{fig:sim_spectrum} and ~\ref{fig:sim_exp_comparison} summarize the simulation results. The Raman soliton shifts more with increasing gas pressure [Fig.~\ref{fig:sim_spectrum}(a)]. Although multiple Raman transitions are included, SSFS generates clean red-shifted solitons. We also observe dispersive waves around \SI{550}{\nm} and residual energy around \SI{1}{\micro\m}. These spectral features are all consistent with the experimental results. The simulated pulse durations are between \num{30} to \SI{41}{\fs}, somewhat shorter than observed experimentally. This results from the uncompensated dispersion from lenses, the sapphire window, the long-pass filter, and the polarizing beam splitter cube. As an example, the spectral evolution of the pulse at \SIadj{60}{\bar} gas pressure is shown in Fig.~\ref{fig:sim_spectrum}(c). At the beginning, several fundamental solitons in a high-order soliton experience different levels of rotational Raman scattering. Eventually the reddest one temporally separates from the rest of the solitons and initiates the smooth spectral red-shift as in Fig.~\ref{fig:sim_spectrum}(c). This reddest Raman soliton is clear in the temporal evolution [Fig.~\ref{fig:sim_spectrum}(d)]. Both the simulated pulse energy and the spectral shift exhibit the same trends as the experiments. The soliton number goes from $1.5$ to $7.5$ when the gas pressure increases, which accounts for the decreasing efficiency with higher pressure (beyond the quantum efficiency of the Raman scattering). However, the simulations predict Raman solitons with energies about \SI{30}{\nano\joule} higher than observed [Fig.~\ref{fig:sim_exp_comparison}(a)], along with soliton wavelengths up to \SI{1800}{\nm} [Fig.~\ref{fig:sim_exp_comparison}(b)]. The simulated spectra also exhibit broader and more-structured dispersive-wave peaks than the experiments. These discrepancies, while not major, are puzzling. We have not found reasonable combinations of parameters that can significantly reduce them. We tentatively attribute them to discrepancies between the actual fiber parameters and the simplified model of fiber propagation constants and loss, which are evident in Fig.~\ref{fig:schematic}(e).

\begin{figure}[h!]
\centering
\includegraphics[width=0.82\linewidth]{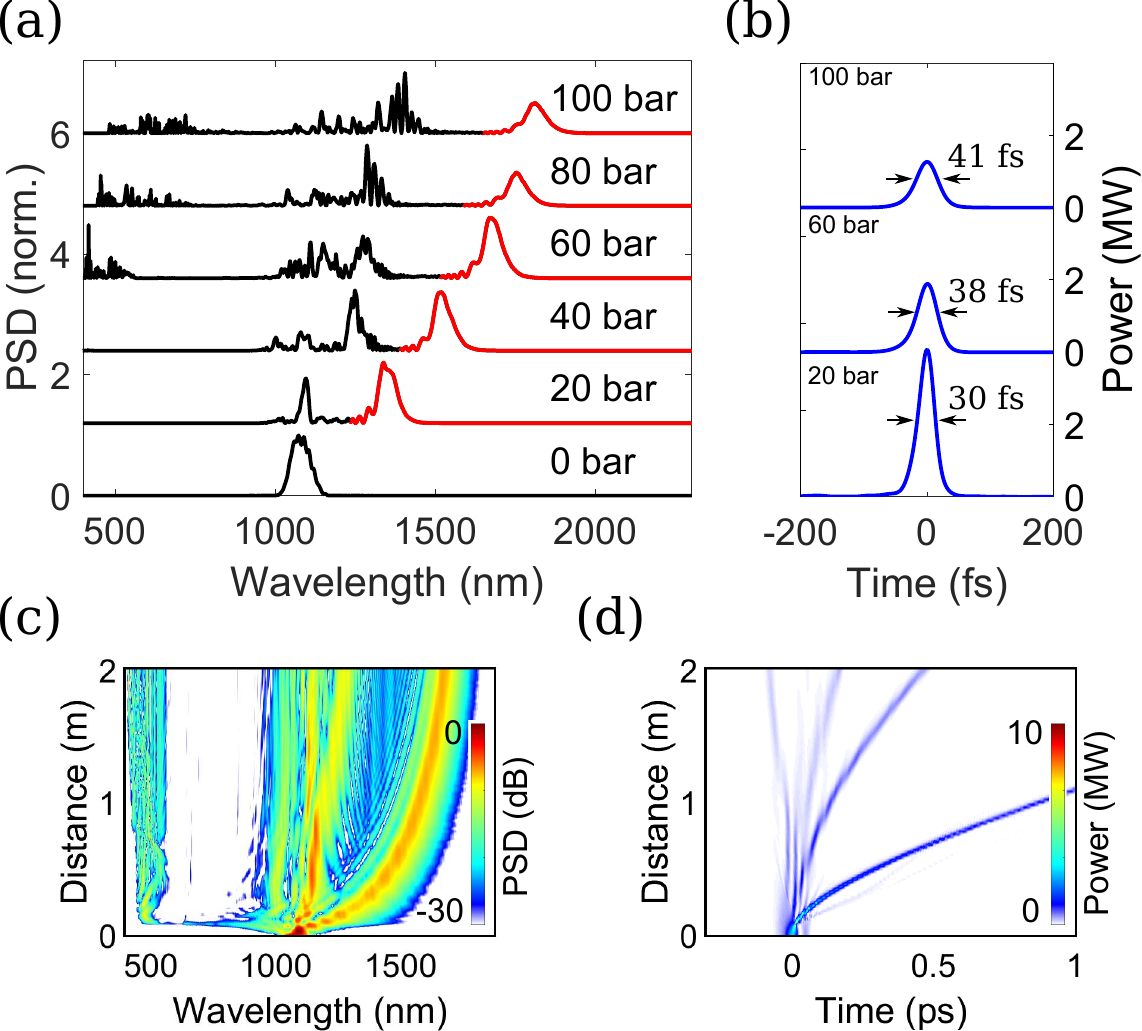}
\caption{Simulated (a) output spectra and (b) Raman pulses at the indicated \ce{H2} pressures. Simulated (c) spectral and (d) temporal evolution at \SIadj{60}{\bar} pressure.}
\label{fig:sim_spectrum}
\end{figure}
\begin{figure}[h!]
\centering
\includegraphics[width=0.82\linewidth]{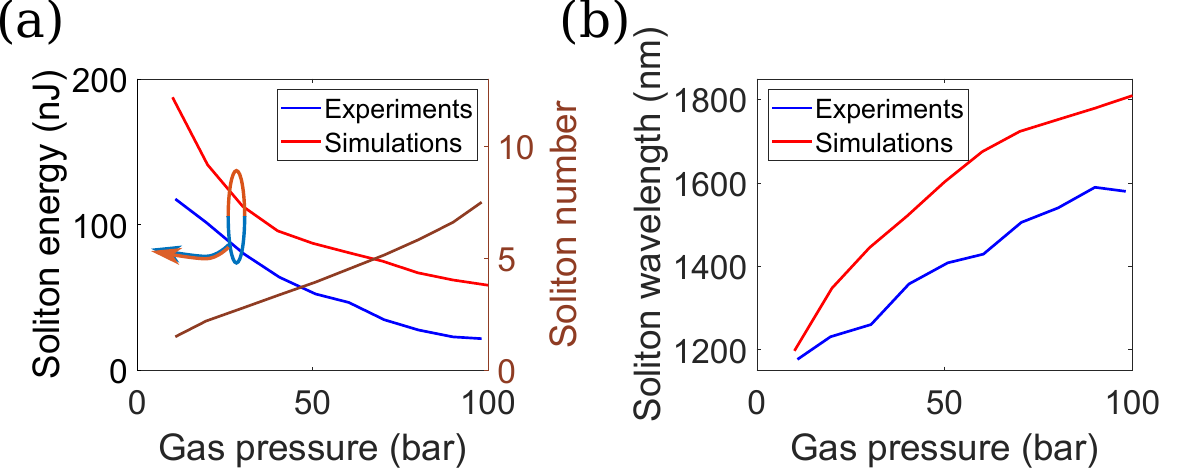}
\caption{(a) Raman soliton energy and soliton number of the input pulse and (b) Raman soliton wavelengths versus gas pressure for both the experiments and the simulations. The launched pulse energy is \SI{240}{\nano\joule}.}
\label{fig:sim_exp_comparison}
\end{figure}

The experimental results presented here are limited by the available input pulse energy. If higher-energy pulses are available, significant increases in the performance should be possible. To assess this, we performed simulations with \SIadj{2}{\micro\joule} and \SIadj{35}{\fs} pulses launched into the same fiber considered above. Such a pulse could be obtained by compression of the output of a standard chirped-pulse amplifier, or by future scaling of the gain-managed concept. The higher pulse energy allows the gas pressure to be reduced. Since the dispersion of a gas-filled hollow-core fiber includes anomalous waveguide dispersion and normal gas dispersion, with the low gas pressure, the dispersion is more anomalous across the spectral range of interest. This, along with weaker nonlinearity, reduces the soliton number. Because generation of dispersive waves is not phase-matched at the lower pressures, no dispersive waves are observed in the output spectra. As a result, the efficiency of Raman-soliton production improves. Wavelengths as long as \SI{1700}{\nm} are generated, with \SIrange[range-phrase = --, range-units = single]{30}{60}{\percent} efficiency (Fig.~\ref{fig:sim_spectrum_2uJ}). Pulse energies above \SIadj{1}{\micro\joule} and peak powers around \SIadj{30}{\MW} are predicted. Ionization of the \ce{H2} should not be a concern at these peak powers; however, it could become significant for scaling to even higher pulse energies (see the discussion of ionization in Supplement 1).

\begin{figure}[h!]
\centering
\includegraphics[width=0.82\linewidth]{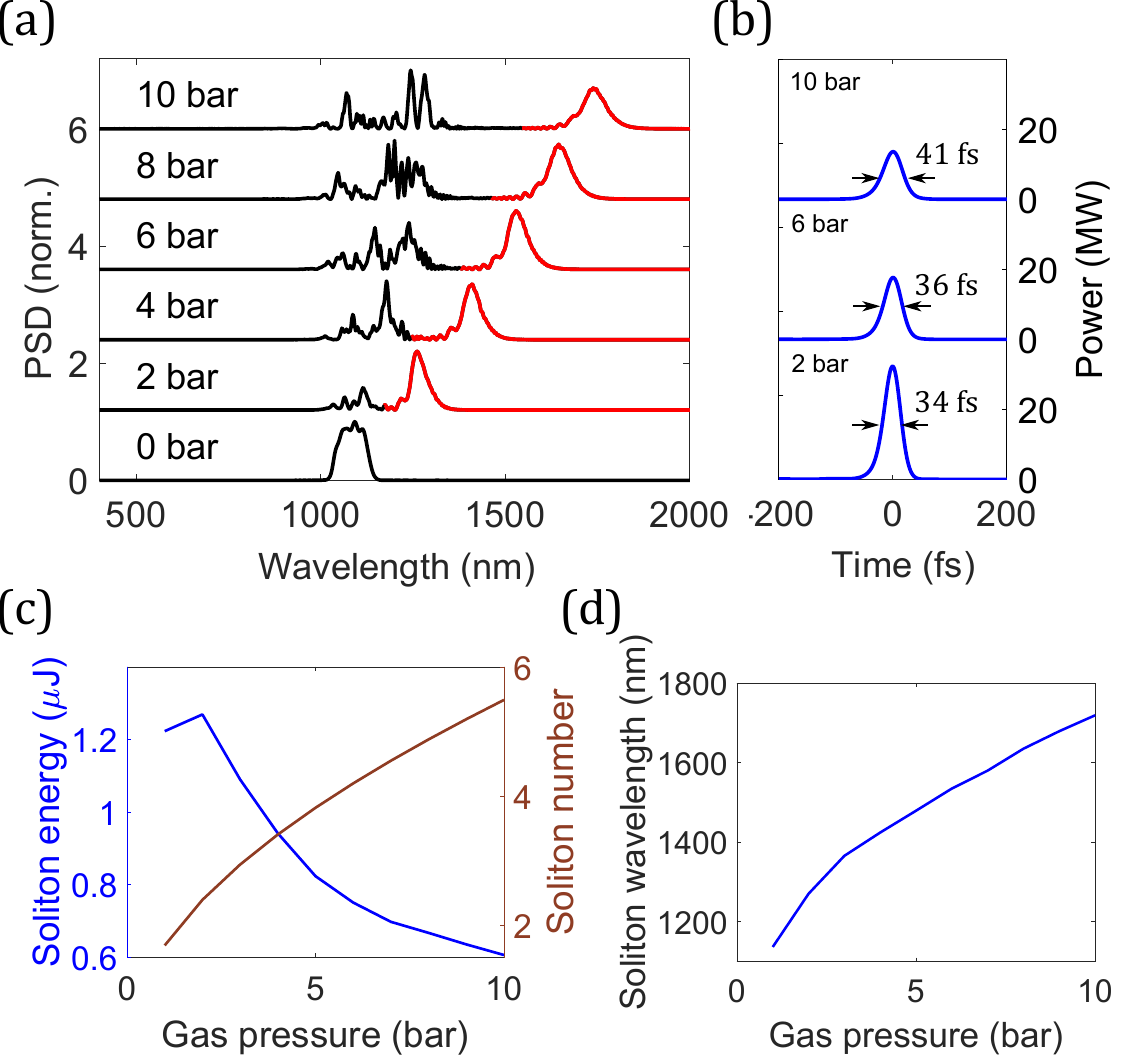}
\caption{Simulations with \SIadj{2}{\micro\joule} and \SIadj{35}{\fs} input pulse. (a) Output spectra and (b) Raman pulses at the indicated \ce{H2} pressures. (c) Raman soliton energy, soliton numbers of the pump pulses, and (d) Raman soliton wavelengths at different gas pressures.}
\label{fig:sim_spectrum_2uJ}
\end{figure}

In conclusion, we have theoretically and experimentally studied SSFS in hydrogen-filled AR-HCF. With short \SIadj{1080}{\nm} input pulses, wavelength tunability from \num{1080} to \SI{1600}{\nm} and consistent \SIadj{45}{\fs} pulse duration are achieved. The generated Raman solitons reach peak powers around \SI{1}{\MW}, and scaling to an order of magnitude higher peak power should be possible.

\begin{backmatter}
\bmsection{Funding} 
Office of Naval Research (N00014-19-1-2592), National Institutes of Health (EB002019), Army Research Office (65016372), and Air Force Research Laboratory (FA86511820019).
\bmsection{Acknowledgments}
The authors are grateful to Francesco Tani for advice on building the gas cells and David Novoa for advice on the simulation code.
\bmsection{Disclosures}
The authors declare no conflicts of interest.
\bmsection{Data availability}
Data underlying the results presented in this paper are not publicly available at this time but may be obtained from the authors upon reasonable request.
\bmsection{Supplemental document}
See Supplement 1 for supporting content.
\end{backmatter}

\clearpage


\title{Efficient soliton self-frequency shift in hydrogen-filled hollow-core fiber: supplemental document}

\begin{abstract}
The supplementary material in this document is organized as follows:
In the manuscript, pulse propagation inside a hollow-core fiber is modeled with a unidirectional pulse propagation equation (UPPE) that includes electronic and (vibrational and rotational) Raman responses. The derivation of the equation is presented in Section 1. Section 2 describes simulations of SSFS in a fiber filled with \ce{N2}. These demonstrate SSFS with lower efficiency due to many Raman transitions that play out equally. Section 3 presents measured output spatial profiles of the filtered Raman solitons generated in the hydrogen-filled fiber, which confirm single-mode propagation. Section 4 discusses, in detail, the observed resonance-induced dispersive-wave generation in Fig.~2 and Fig.~3. Section 5 is a brief discussion of the possible role of photoionization in future work on scaling results presented here to higher pulse energies.
\end{abstract}

\section{UPPE with delayed Raman response of a gas}

\subsection{UPPE}

The convention of the field follows the traditional generalized nonlinear Schr\"{o}dinger equation (GNLSE) \cite{Poletti2008}:
\begin{align}
\vec{\mathbb{E}}(\vec{x},t) & =\frac{1}{2}\left[\vec{\mathcal{E}}(\vec{x},t)+\text{c.c.}\right] \nonumber \\
& =\int\diff\omega\frac{1}{2}\left\{\frac{\vec{F}(x,y,\omega)}{N(\omega)}A(z,\omega)\E^{i\left[\beta(\omega)z-\omega t\right]}+\text{c.c.}\right\} \nonumber \\
& =\frac{1}{2}\left\{\frac{\vec{F}(x,y)}{N}\left[A(z,t)\E^{i\left(\beta_0z-\omega_0t\right)}\right]+\text{c.c.}\right\}\text{, assume }\vec{F}(x,y,\omega)=\vec{F}(x,y) \nonumber \\
& =\frac{1}{2}\left\{\frac{\vec{F}(x,y)}{\sqrt{\frac{\epsilon_0n_{\text{eff}}c}{2}}}\left[A(z,t)\E^{i\left(\beta_0z-\omega_0t\right)}\right]+\text{c.c.}\right\}, \label{eq:E_in_W}
\end{align}
where $\vec{\mathcal{E}}(\vec{x},t)$ is the analytic signal of the real-valued electric field $\vec{\mathbb{E}}(\vec{x},t)$, ``c.c.'' stands for complex conjugate, $\vec{F}(x,y)$ is the normalized spatial mode profile of the fundamental mode with the normalization condition, $\int\abs{\vec{F}}^2\diff^2x=1$, and is assumed to be independent of frequency. $\vec{\mathbb{E}}(\vec{x},t)$ has the unit of \si{\volt/\meter}. $A(z,t)$ represents the field and is normalized to have the unit of \si{\sqrt{\watt}} with the normalization constant $N=\sqrt{\frac{\epsilon_0n_{\text{eff}}c}{2}}$. $\beta(\omega)=n_{\text{eff}}(\omega)k_0$ is the propagation constant. $\beta_0$ and $\omega_0$ are two free parameters, usually chosen as the propagation constant and the angular frequency at the center frequency of a pulse. $A$ has the following time-frequency relation,
\begin{equation}
A(z,t)=\int\diff\omega A(z,\omega)\E^{i\left[\left(\beta(\omega)-\beta_0\right)z-\left(\omega-\omega_0\right)t\right]}
\end{equation}

The scalar-field UPPE is
\begin{equation}
\partial_zA(z,\Omega)=i\left[\beta(\omega)-\left(\beta_0+\beta_1\Omega\right)\right]A(z,\Omega)+\frac{i\omega}{4N^2}P(z,\Omega),
\end{equation}
where $\beta_1$ is the inverse group velocity of the moving frame, the analytic signal of the nonlinear polarization $\vec{\mathcal{P}}(\vec{x},t)=\frac{\vec{F}(x,y)}{N}P(z,t)\E^{i\left(\beta_0z-\omega_0t\right)}$ and $\Omega=\omega-\omega_0$. By expanding $P(z,\Omega)$ into Kerr and Raman terms for the linearly-polarized field, it becomes
\begin{align}
\partial_zA(z,\Omega)=i&\left[\beta(\omega)-\left(\beta_0+\beta_1\Omega\right)\right]A(z,\Omega)+ \nonumber \\
& \frac{i\omega}{\epsilon_0^2n_{\text{eff}}^2c^2A_{\text{eff}}(\omega)}\left(\frac{3\epsilon_0\chi^{(3)}_{\text{electronic}}}{4}\mathfrak{F}[\abs{A}^2A]+\mathfrak{F}\left[A\left[R(t)\ast \left(\abs{A}^2\right)\right]\right]\right),
\label{eq:UPPE}
\end{align}
where $R(t)=R^{\text{rot}}(t)+R^{\text{vib}}(t)$, the total Raman response including rotational and vibrational parts. $\mathfrak{F}$ stands for the Fourier Transform and $\ast$ for the convolution operation \cite{Agrawal2013}.

\subsection{Rotational Raman}

We modified the model of the rotational Raman response introduced by Chen \etal \cite{Chen2007} and Wahlstrand \etal \cite{Wahlstrand2015} to obtain a delayed Raman response, $R(t)$, of the form commonly used for silica fibers. Their model has been applied in several studies; however, instead of calculating the analytical form of the Raman response, they fit the calculated molecular orientation $\expval{\cos^2\theta}_t$ to the Raman response function modeled with a damped harmonic oscillator $R(t)=R_0\E^{-\gamma_Rt}\sin\left(\omega_Rt\right)$ where $R_0$ is the normalization constant such that $\int R(t)\diff t=1$ \cite{Langevin2019,Fan2020,Carpeggiani2020,Safaei2020,Beetar2020}. Here, we derive an analytical expression for the rotational Raman response and show that it can be implemented in the UPPE [Eq.~(\ref{eq:UPPE})] directly. By doing so, long-time features of the long dephasing times of gases, such as the \SIadj{2.1}{\ps} revivals of coherence or molecular re-alignment in \ce{N2} \cite{Nibbering1997,Bustard2008} or the \SIadj{100}{\ps} ringing of the coherence wave in \ce{H2}, can be captured.

We start with the dielectric response of diatomic gas molecules

\begin{align}
\epsilon & =\epsilon_0+N_g\expval{\alpha}_t \nonumber \\
& =\epsilon_0+N_g\left(\triangle\alpha\expval{\cos^2\theta}_t+\alpha^{\perp}\right) \nonumber \\
& =\epsilon(t\rightarrow-\infty)+N_g\triangle\alpha\left(\expval{\cos^2\theta}_t-\frac{1}{3}\right),
\end{align}
where $N_g$ is the molecular number density, $\triangle\alpha=\alpha_{\parallel}-\alpha_{\perp}$ is the polarizability anisotropy. $\alpha_{\parallel}$ and $\alpha_{\perp}$ are molecular polarizabilities when the electric field is parallel and perpendicular to the molecule, respectively. And
\begin{equation}
\epsilon(t\rightarrow-\infty)=\epsilon_0+N_g\left(\frac{\triangle\alpha}{3}+\alpha^{\perp}\right).
\end{equation}

To solve $\expval{\cos^2\theta}_t$, the density-matrix approach is applied.
\begin{equation}
\expval{\cos^2\theta}_t=\Tr\left[\hat{\rho}\cos^2\theta\right]
\label{eq:cos2}
\end{equation}
Here $\cos^2\theta$ is treated as an operator and has $\left(\cos^2\theta\right)_{kl}=\mel{k}{\cos^2\theta}{l}$. $\hat{\rho}$ is the density matrix. From the perturbation theory \cite{Boyd2008}, $\hat{\rho}=\hat{\rho}^{(0)}+\hat{\rho}^{(1)}$ where
\begin{equation}
\rho^{(1)}_{kl}=-\frac{i}{\hbar}\int^t_{-\infty}\diff\tau\left[H_{\text{int}}(\tau),\hat{\rho}^{(0)}\right]_{kl}\E^{\left(\gamma_{kl}+i\omega_{kl}\right)\left(\tau-t\right)}
\end{equation}
is the first-order correction to the density matrix induced by the perturbed Hamiltonian,
\begin{align}
H_{\text{int}} & =\expval{-\int\vec{\mathbb{E}}\cdot\diff\vec{\mu}}_t \nonumber \\
& =\expval{-\sum_{i,j}\alpha_{ij}\int\mathbb{E}^i\cdot\diff\mathbb{E}^j}_t=-\expval{\sum_{i,j}e^i\alpha_{ij}e^j}_t\int\abs{\vec{\mathbb{E}}}\cdot\diff\abs{\vec{\mathbb{E}}}=-\frac{1}{2}\left[\triangle\alpha\left(\cos^2\theta\right)+\alpha^{\perp}\right]\abs{\vec{\mathbb{E}}}^2, 
\label{eq:Hint}
\end{align}
where $\vec{\mu}=\rttensor{\alpha}\cdot\vec{\mathbb{E}}$.

As for the zeroth-order density matrix $\rho^{(0)}_{J,M}$,
\begin{subequations}
\begin{align}
& \rho^{(0)}_{J,M}=\frac{g_J\E^{-\frac{E_J}{k_BT}}}{Z} \\
& Z=\sum_Jg_J(2J+1)\E^{-\frac{E_J}{k_BT}} \\
& E_J=B_eJ(J+1)-D_eJ^2(J+1)^2,
\end{align}
\end{subequations}
where $g_J$ is the nuclear-spin statistical factor, and $B_e$ and $D_e$ are constants for the rotational energy states.

\begin{align}
\left[H_{\text{int}}(\tau),\hat{\rho}^{(0)}\right]_{kl} & =\left(\rho^{(0)}_l-\rho^{(0)}_k\right)H_{\text{int},kl},\quad\because\rho^{(0)}_{kl}=\rho^{(0)}_k\delta_{kl}\text{ when there's no external field} \nonumber \\
& =\left(\rho^{(0)}_l-\rho^{(0)}_k\right)\frac{1}{2}\left[-\triangle\alpha\left(\cos^2\theta\right)_{kl}-\alpha^{\perp}\delta_{kl}\right]\abs{\vec{\mathbb{E}}}^2
\end{align}

Because the rotational eigenstates $\ket{k}=\ket{J,M}=Y_{JM}(\theta,\phi)$ are spherical harmonics, 
\begin{equation}
\left(\cos^2\theta\right)_{kl}=\mel{J,M}{\cos^2\theta}{J',M'}
\end{equation}
is nonvanishing only for $M=M'$ and $J=J',J'\pm2$. Therefore, we obtain
\begin{equation}
\rho^{(1)}_{J+2,J,M}=\frac{i}{2\hbar}\left(\rho^{(0)}_{J,M}-\rho^{(0)}_{J+2,M}\right)\triangle\alpha\left(\cos^2\theta\right)_{J+2,J}^M\left[\E^{\left(-\gamma_{J+2,J}-i\omega_{J+2,J}\right)t}\ast\abs{\vec{\mathbb{E}}}^2\right]
\end{equation}
Note that $\rho^{(1)}_{J,J+2,M}$ is implicitly considered as well since 
\begin{equation}
\rho^{(1)}_{J,J+2,M}=\left(\rho^{(1)}_{J+2,J,M}\right)^{\ast}\text{ and }\left[\left(\cos^2\theta\right)_{J,J+2}^M\right]^{\ast}=\left(\cos^2\theta\right)_{J+2,J}^M.
\label{eq:rho_JJ+2}
\end{equation}
The integration range in the convolution $\int^{\infty}_{-\infty}$ can be transformed into $\int^t_{-\infty}$ by assuming its integrand is defined only within the range of $[0,t]$ which is true for pulses.

Eq.~(\ref{eq:rho_JJ+2}) is then incorporated into Eq.~(\ref{eq:cos2}) and becomes
\begin{align}
\expval{\cos^2\theta}_t & =\Tr\left[\hat{\rho}\cos^2\theta\right] \nonumber \\
& =\sum_{kl}\left(\rho^{(0)}_{kl}+\rho^{(1)}_{kl}\right)\left(\cos^2\theta\right)_{lk} \nonumber \\
& =\sum_{k}\rho^{(0)}_{k}\left(\cos^2\theta\right)_{kk}+\sum_{kl}\rho^{(1)}_{kl}\left(\cos^2\theta\right)_{lk}\quad\because\rho^{(0)}_{kl}=\rho^{(0)}_{k}\delta_{kl} \nonumber \\
& =\frac{1}{3}+\sum_{JM}\left(\rho^{(1)}_{J+2,J,M}+\rho^{(1)}_{J,J+2,M}\right)\left(\cos^2\theta\right)_{J+2,J}^M\quad\because\left(\cos^2\theta\right)_{J+2,J}^M\text{ is real} \nonumber \\
& =\frac{1}{3}+\sum_{JM}2\Re\left[\rho^{(1)}_{J+2,J,M}\right]\left(\cos^2\theta\right)_{J+2,J}^M
\end{align}

We then obtain 
\begin{align}
\epsilon & =\epsilon(t\rightarrow-\infty)+\sum_{JM}2N_g\triangle\alpha\Re\left[\rho^{(1)}_{J+2,J,M}\right]\left[\left(\cos^2\theta\right)_{J+2,J}^M\right]^2 \nonumber \\
& =\epsilon(t\rightarrow-\infty)-\sum_{JM}\frac{1}{\hbar}N_g\left(\triangle\alpha\right)^2\left(\rho^{(0)}_{J,M}-\rho^{(0)}_{J+2,M}\right)\left[\left(\cos^2\theta\right)_{J+2,J}^M\right]^2\Im\left[\E^{\left(-\gamma_{J+2,J}-i\omega_{J+2,J}\right)t}\ast\abs{\vec{\mathbb{E}}}^2\right] \nonumber \\
& =\epsilon(t\rightarrow-\infty)+\sum_{JM}\frac{1}{\hbar}N_g\left(\triangle\alpha\right)^2\left(\rho^{(0)}_{J,M}-\rho^{(0)}_{J+2,M}\right)\left[\left(\cos^2\theta\right)_{J+2,J}^M\right]^2\Im\left[\E^{\left(-\gamma_{J+2,J}+i\omega_{J+2,J}\right)t}\ast\abs{\vec{\mathbb{E}}}^2\right]\label{eq:epsilon_Im}
\end{align}

By applying \cite{Chen2007}
\begin{align}
\sum_{M=-J}^J\left[\left(\cos^2\theta\right)_{J+2,J}^M\right]^2 & =\sum_{M=-J}^J\frac{\left[\left(J+2\right)^2-M^2\right]\left[\left(J+1\right)^2-M^2\right]}{\left(2J+1\right)\left(2J+3\right)^2\left(2J+5\right)} \nonumber \\
& =\frac{2}{15}\frac{(J+2)(J+1)}{2J+3},
\end{align}
Eq.~(\ref{eq:epsilon_Im}) becomes
\begin{equation}
\epsilon=\epsilon(t\rightarrow-\infty)+\sum_J\frac{2}{15\hbar}N_g\left(\triangle\alpha\right)^2\left(\rho^{(0)}_J-\rho^{(0)}_{J+2}\right)\frac{(J+2)(J+1)}{2J+3}\Im\left[\E^{\left(-\gamma_{J+2,J}+i\omega_{J+2,J}\right)t}\ast\abs{\vec{\mathbb{E}}}^2\right],
\label{eq:epsilon_rot}
\end{equation}
where $\hat{\rho}^{(0)}$ is independent of $M$: $\rho^{(0)}_J=\rho^{(0)}_{J,M}$.

If we ignore the highly-oscillatory term in $\abs{\vec{\mathbb{E}}}^2$, we can approximate $\abs{\vec{\mathbb{E}}}^2$ as $\abs{\vec{\mathcal{E}}}^2/2$.

Because $\vec{\mathbb{P}}=\triangle\epsilon\vec{\mathbb{E}}=\left[R(t)\ast\abs{\vec{\mathcal{E}}}^2\right]\vec{\mathbb{E}}$, we can then obtain the Raman response function,
\begin{align}
R^{\text{rot}} & =N_g\sum_J\frac{1}{15\hbar}\left(\triangle\alpha\right)^2_J\left(\rho^{(0)}_J-\rho^{(0)}_{J+2}\right)\frac{(J+2)(J+1)}{2J+3}\E^{-\gamma^{0J\to0J+2}_2t}\sin\left(\omega_{0J+2,0J}t\right) \nonumber \\
& =N_g\frac{1}{15\hbar}\left(\triangle\alpha\right)^2\E^{-\gamma_2^{\text{rot}}t}\sum_J\left(\rho^{(0)}_J-\rho^{(0)}_{J+2}\right)\frac{(J+2)(J+1)}{2J+3}\sin\left(\omega_{0J+2,0J}t\right),
\label{eq:R_rot}
\end{align}
where all dephasing times $T_2^{\text{rot}}=\frac{1}{\gamma_2^{\text{rot}}}=\frac{1}{\gamma_2^{0J\to0J+2}}$ are assumed to be the same.

\subsection{Vibrational Raman}

Although Wahlstrand \etal provide a detailed derivation \cite{Wahlstrand2015},  we derive the equation with a different approach that starts from the Maxwell-Bloch equations \cite{Kalosha2000,Kien2002,Belli2015}.

\begin{subequations}
\begin{align}
\dot{w} & =-\gamma_1\left(w+1\right)-\frac{2[\alpha]_{ab}}{\hbar}\Im{\rho_{ab}}\abs{\vec{\mathbb{E}}(t)}^2 \label{eq:wdot} \\
\dot{\rho}_{ab} & =\left(-\gamma_2+i\omega_{ba}\right)\rho_{ab}+\frac{i}{2\hbar}\left[\left([\alpha]_{aa}-[\alpha]_{bb}\right)\rho_{ab}+[\alpha]_{ab}w\right]\abs{\vec{\mathbb{E}}(t)}^2, \label{eq:rhodot}
\end{align}
\end{subequations}
where $[\rho]$ is the density matrix so that $w=\rho_{bb}-\rho_{aa}$ is the population inversion, $\gamma_1$ and $\gamma_2$ are the dephasing time of the coherence and the decay time of the upper-state population, respectively, $[\alpha]$ is the molecular polarizability matrix.

We assume that the applied field only perturbs the system such that
\begin{subequations}
\begin{align}
& \rho_{bb}\approx0\quad\Rightarrow\quad w\approx-1 \\
& \rho_{ab}\approx0.
\end{align}
\end{subequations}
Eq.~(\ref{eq:rhodot}) then can be solved by use of an integrating factor:
\begin{equation}
\rho_{ab}(t)=-\frac{i}{2\hbar}[\alpha]_{ab}\left[\E^{\left(-\gamma_2+i\omega_{ba}\right)t}\ast\abs{\vec{\mathbb{E}}(t)}^2\right]
\end{equation}

The Raman polarization 
\begin{align}
\vec{\mathbb{P}}_R & =N_g\expval{\hat{\alpha}\vec{\mathbb{E}}}=N_g\left(\hat{\rho}\hat{\alpha}\right)\vec{\mathbb{E}} \nonumber \\
& =N_g\left[[\alpha]_{aa}+\left([\alpha]_{bb}-[\alpha]_{aa}\right)\rho_{bb}+2[\alpha]_{ab}\Re\left\{\rho_{ab}\right\}\right]\vec{\mathbb{E}},
\label{eq:PR}
\end{align}
in which $N_g[\alpha]_{aa}$ is the linear term and is taken into account in the propagation constant $\beta(\omega)$. Hence, the nonlinear Raman polarization is
\begin{align}
\vec{\mathbb{P}}_{R\text{, nonlinear}} & =2N_g[\alpha]_{ab}\Re\left\{\rho_{ab}\right\}\vec{\mathbb{E}} \nonumber \\
& =\frac{1}{\hbar}N_g[\alpha]_{ab}^2\left[\E^{-\gamma_2t}\sin\left(\omega_{ba}t\right)\ast\abs{\vec{\mathbb{E}}(t)}^2\right]
\label{eq:Pvib_alpha_eff}
\end{align}

By comparing the nonlinear Raman polarization in \cite{Boyd2008} (along with careful treatments of analytic signals) with Eq.~(\ref{eq:Pvib_alpha_eff}), we derive the following relation
\begin{equation}
[\alpha]_{ab}=\sqrt{\frac{\hbar}{2\mu\omega_{ba}}}\left(\dod{\alpha}{\mathbb{Q}}\right)_0
\label{eq:alpha_eff_Da}
\end{equation}
where $\mu$ is the reduced mass of the gas molecule, $\mathbb{Q}$ is its normal coordinate, and the subscript $0$ denotes that the derivative is taken at equilibrium. With it, we obtain
\begin{equation}
\vec{\mathbb{P}}_{R\text{, nonlinear}}=\frac{1}{2\mu\omega_{ba}}N_g\left(\dod{\alpha}{\mathbb{Q}}\right)_0^2\left[\E^{-\gamma_2t}\sin\left(\omega_{ba}t\right)\ast\abs{\vec{\mathbb{E}}(t)}^2\right]
\label{eq:Pvib}
\end{equation}

Next we consider the population of each state and add one more term corresponding to different rotational energy states \cite{Wahlstrand2015}. Eq.~(\ref{eq:Pvib}) becomes
\begin{multline}
\vec{\mathbb{P}}_{R\text{, nonlinear}}=\sum_JN_g\frac{(2J+1)\rho^{(0)}_J}{2\mu\omega_{1J,0J}}\left[\left(\dod{\alpha}{\mathbb{Q}}\right)^2_0+\frac{4}{45}\frac{J(J+1)}{(2J-1)(2J+3)}\left(\dod{\left(\triangle\alpha\right)}{\mathbb{Q}}\right)^2_0\right]\times \\
\left[\E^{-\gamma_2t}\sin\left(\omega_{1J,0J}t\right)\ast\abs{\vec{\mathbb{E}}(t)}^2\right]\vec{\mathbb{E}}
\label{eq:PRJ}
\end{multline}

With $\abs{\vec{\mathbb{E}}}^2\approx\abs{\vec{\mathcal{E}}}^2/2$ and $\vec{\mathbb{P}}=\left[R(t)\ast\abs{\vec{\mathcal{E}}}^2\right]\vec{\mathbb{E}}$, we have the vibrational Raman response for the Q branch,
\begin{equation}
R^{\text{vib}}=N_g\frac{1}{4\mu}\E^{-\gamma_2^{\text{vib}}t}\sum_J(2J+1)\rho^{(0)}_J\frac{\left(\od{\alpha}{\mathbb{Q}}\right)^2_0+\frac{4}{45}\frac{J(J+1)}{(2J-1)(2J+3)}\left(\od{\left(\triangle\alpha\right)}{\mathbb{Q}}\right)^2_0}{\omega_{1J,0J}}\sin\left(\omega_{1J,0J}t\right)
\label{eq:R_vib}
\end{equation}

\subsection{Required steps in numerical simulations}

Due to long dephasing times of the Raman responses, the operation of the convolution ($\ast$) can't be performed directly with the circular convolution theorem unless the time window in simulations is large enough to cover the entire Raman response which is typically on the order of \num{1} to \SI{10}{\ns}. This is undesirable especially if the physical phenomenon of interest happens on the time scale of less than a few picoseconds. Therefore, in each nonlinear step of ERK4(3)-IP \cite{Balac2013}, doubling the time window by filling one side with dummy zeros is required to apply circular convolution theorem and avoids aliasing numerically. Only the signal in the original time window contains useful information and is recovered back after the nonlinear operation.

Because discrete Raman transitions generate signals far from the pump frequency, aliasing occurs if the frequency window isn't large enough. Due to various orders of Raman transitions, aliasing is difficult to avoid no matter how large the frequency window is. Hence, in each nonlinear step, we extend the frequency window by three times by filling both edges with zeros. After the computation of the Raman response, we downsample the frequency window back to its original size and discard the high and low frequency parts. This operation avoids the potential aliasing resulting from Raman transitions of various orders and guarantees that the computation considers only the physics within the desired frequency window.

\subsection{Parameter values}

In tables ~\ref{tab:n} to \ref{tab:T2}, the parameter values for both \ce{H2} and \ce{N2} are listed. \ce{N2} has been commonly used in  hollow-core fibers. It will be used as an example to explain why \ce{H2} is advantageous for efficient Raman-soliton formation.

\begin{table}[h!]
    \centering
    \caption{\bf Refractive index $n_0$ at \SI{0}{\degreeCelsius} and \SI{1}{\bar}. At temperature $T$, $n=1+\left(n_0-1\right)\rho$, where $\rho$ is the gas density in amagats. $\lambda$ is in \si{\micro\meter}.}
    \begin{tabular}{c|l}
        \toprule
        Molecule & $n_0$ \\
        \midrule
        \ce{H2} & $1+\frac{0.0148956}{180.7-\lambda^{-2}}+\frac{0.0049037}{92-\lambda^{-2}}$ \cite{Peck1977} \\
        \ce{N2} & $1+\num{6.8552e-5}+\frac{\num{3.243157e-2}}{144-\lambda^{-2}}$ \cite{Peck1966} \\
        \bottomrule
    \end{tabular}
    \label{tab:n}
\end{table}
\begin{table}[h!]
    \centering
    \caption{\bf Nonlinear refractive index $n_2=n_2^0\rho$, where $\rho$ is the gas density in amagats. For \ce{N2}, $n_2^0=10^{24}\frac{P_{n_2}^{-1}}{\lambda_0^{-2}-\lambda^{-2}}$ ($10^{-24}$ \si{\m^2/\W/\bar}) \cite{Brown2018}.}
    \begin{tabular}{c|ccc}
        \toprule
        Molecule & $n_2^0$ ($10^{-24}$ \si{\m^2/\W/\bar}) & $P_{n_2}$ (\si{\W}) & $\lambda_0$ (\si{\m}) \\
        \midrule
        \ce{H2} & 6.5 \cite{Koehler2013,Wahlstrand2015} && \\
        \ce{N2} && \num{14.63e9} & \num{0.3334e-6} \\
        \bottomrule
    \end{tabular}
    \label{tab:n2}
\end{table}
\begin{table}[h!]
    \centering
    \caption{\bf Nuclear spin statistics $g_J$ \cite{MIT:Physical_Chemistry}}
    \begin{tabular}{c|cc}
        \toprule
        Molecule & $J$ is even & $J$ is odd \\
        \midrule
        \ce{H2} & 1 & 3 \\
        \ce{N2} & 6 & 3 \\
        \bottomrule
    \end{tabular}
    \label{tab:g_J}
\end{table}
\begin{table}[h!]
    \centering
    \caption{\bf Constants of energy states \cite{Spelsberg1994,Demtroeder2010,Lampel2015}}
    \begin{tabular}{c|cccc}
        \toprule
        Molecule & $B_e$ (\si{cm^{-1}}) & $D_e$ (\si{cm^{-1}}) & $\omega_{\text{vib}}$ (\si{cm^{-1}}) \\
        \midrule
        \ce{H2} & 60.8 & \num{1.6e-2} & 4155 \cite{Weber1994} \\
        \ce{N2} & 1.98958 & \num{5.76e-6} & 2329.9 \\
        \bottomrule
    \end{tabular}
    \label{tab:BD}
\end{table}
\begin{table}[h!]
    \centering
    \caption{\bf Polarizability. $\si{\text{a.u.}}=\SI{1.64878e-41}{Fm^2}$. $\mu$ is the reduced mass of the molecule. Their values are carefully adjusted to obtain agreements with other studies \cite{Chen2007,Beetar2020,Li2020,Konyashchenko2008,Bischel1986a}.}
    \begin{tabular}{c|ccc}
        \toprule
        Molecule & $\triangle\alpha$ (a.u.) & $\alpha'/\sqrt{\mu}$ (\si{\F\m/\sqrt{\kg}}) & $\left(\triangle\alpha\right)'/\sqrt{\mu}$ (\si{\F\m/\sqrt{\kg}}) \\
        \midrule
        \ce{H2} & 2.23 \cite{Bridge1966,Kolos2020} & \num{3.88e-17} \cite{Wahlstrand2015} & \num{2.82e-17} \cite{Kolos2020} \\
        \ce{N2} & 4.54 \cite{Langevin2019,Bridge1966} & \num{1.80e-17} \cite{Lampel2015} & \num{2.29e-17} \cite{Lampel2015} \\
        \bottomrule
    \end{tabular}
    \label{tab:alpha}
\end{table}
\begin{table}[h!]
    \centering
    \caption{\bf Dephasing time $T_2=\frac{1}{\pi\triangle\nu}=\frac{1}{\gamma_2}$. $\rho$ is the gas density in amagats and $T$ is the temperature in \si{\K}. Below $\triangle\nu$ is under \si{\MHz}.}
    \begin{tabular}{c|cc}
        \toprule
        Molecule & vibrational Raman $\triangle\nu^{\text{vib}}$ & rotational Raman $\triangle\nu^{\text{rot}}$ \\
        \midrule
        \ce{H2} & {\scriptsize$\frac{309}{\rho}\left[\frac{T}{298}\right]^{0.92}+\left[51.8+0.152\left(T-298\right)+4.85\times10^{-4}\left(T-298\right)^2\right]\rho$ \cite{Bischel1986}} & $\frac{6.15}{\rho}+114\rho$ \cite{Weber1994} \\
        \ce{N2} & $22.5,~\rho<10$ \cite{Weber1994} & $3570\rho$ \cite{Weber1994} \\
        \bottomrule
    \end{tabular}
    \label{tab:T2}
\end{table}
\newpage

\section{Simulations of pulse propagation in nitrogen-filled hollow-core fiber}

To demonstrate what will happen if there are potentially many Raman transitions, we consider nitrogen-filled hollow-core fiber. Because \ce{N2} has closely-spaced rotational energy states, the pulse is affected by multiple Raman transitions simultaneously. To illuminate the underlying physics and compare \ce{H2} and \ce{N2}, we simulated injection of a \SIadj{20}{\fs} fundamental soliton into a \SIadj{30}{\micro\meter}-core-diameter AR-HCF with \SI{10}{\bar} of \ce{H2} or \ce{N2}. Under this gas pressure, the dispersion is dominated by the anomalous waveguide dispersion. The fundamental soliton energies are \SI{220}{\nano\joule} for \ce{H2} and \SI{180}{\nano\joule} for \ce{N2}. A pulse duration of \SI{20}{\fs} is chosen to obtain a strong fundamental soliton and broad bandwidth for clear demonstration of SSFS.

\begin{figure}[h!]
\centering
\includegraphics[width=0.9\linewidth]{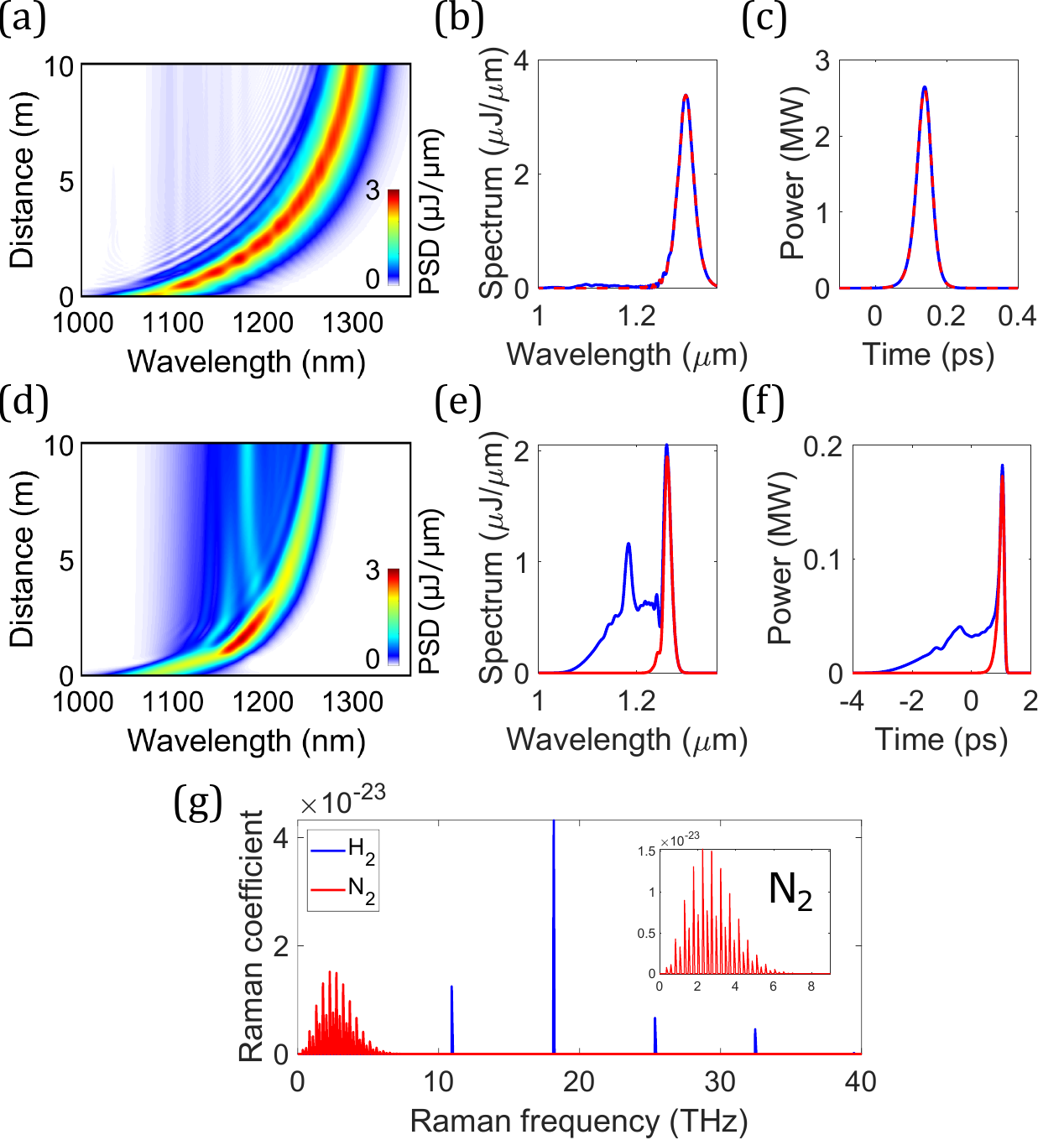}
\caption{Simulations with a fundamental soliton. Simulated (a) spectral evolution, (b) output spectrum, and (c) output pulse before (blue) and after (red) filtering out the reddest spectral peak in \ce{H2}. Red dashed lines are used to clearly show their overlap. Simulated (d) spectral evolution, (e) output spectrum, and (f) output pulse before (blue) and after (red) filtering out the reddest spectral peak in \ce{N2}. (g) The strength of each Raman transition. It's calculated under the SI unit with $A(z,t)$ having the unit of \si{\sqrt{\watt}}. The inset in (g) is a close-view of the \ce{N2} Raman strength.}
\label{fig:N2_soliton}
\end{figure}

Figs.~\ref{fig:N2_soliton}(a-c) show the SSFS in \ce{H2} while Figs.~\ref{fig:N2_soliton}(d-f) show it in \ce{N2}. In contrast to the single clean $\sech$-like Raman soliton in \ce{H2}, there is a wide spectrum in \ce{N2} [Fig.~\ref{fig:N2_soliton}(b) and \ref{fig:N2_soliton}(e)]. In Fig.~\ref{fig:N2_soliton}(d), at \SIadj{1}{\m} and \SIadj{2}{\m} propagation distances, the Raman soliton develops blue spectral peaks. This results from the interplay of various Raman transitions in \ce{N2}. If we filter out only the reddest spectral peak, we can see, in Fig.~\ref{fig:N2_soliton}(c), all the energy contributes to the \SIadj{45}{\fs} Raman soliton. However, in Fig.~\ref{fig:N2_soliton}(f), the pulse develops a long pesdestal while the filtered pulse takes up only a fraction of the energy and has \SIadj{80}{\fs} pulse duration. To demonstrate the difference between the Raman transitions in both gases, we define the strength of the transition from the rotational energy state $J$ to $J+2$ as the $R^{\text{coeff}}_J$ in Eq.~(\ref{eq:R_coeff}), which is derived from Eq.~(\ref{eq:R_rot}). In Fig.~\ref{fig:N2_soliton}(g), we can see that there is only one dominant Raman transition in \ce{H2} around \SI{18}{\THz} which results from S(1). On the other hand, there is a cluster of transitions with similar strengths at \SI{3}{\THz} in \ce{N2}.

\begin{equation}
R^{\text{rot}}=\E^{-\gamma_2^{\text{rot}}t}\sum_JR^{\text{coeff}}_J\sin\left(\omega_{0J+2,0J}t\right)
\label{eq:R_coeff}
\end{equation}

\newpage

\section{Measurements of the output spatial profiles}

\begin{figure}[h!]
\centering
\includegraphics[width=\linewidth]{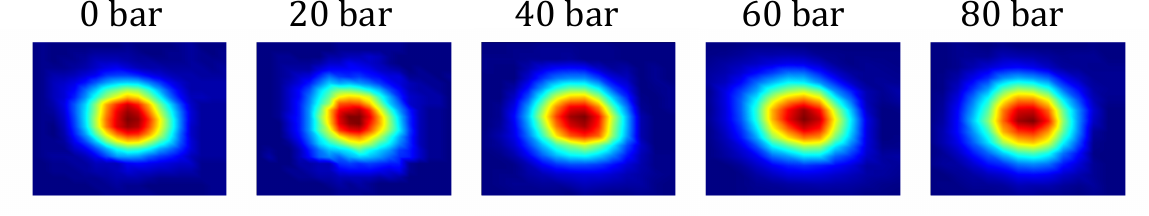}
\caption{Measured output spatial profiles with \ce{H2} pressure from \SI{0}{\bar} to \SI{80}{\bar}.}
\label{fig:output_profile}
\end{figure}

\begin{figure}[h!]
\centering
\includegraphics[width=0.7\linewidth]{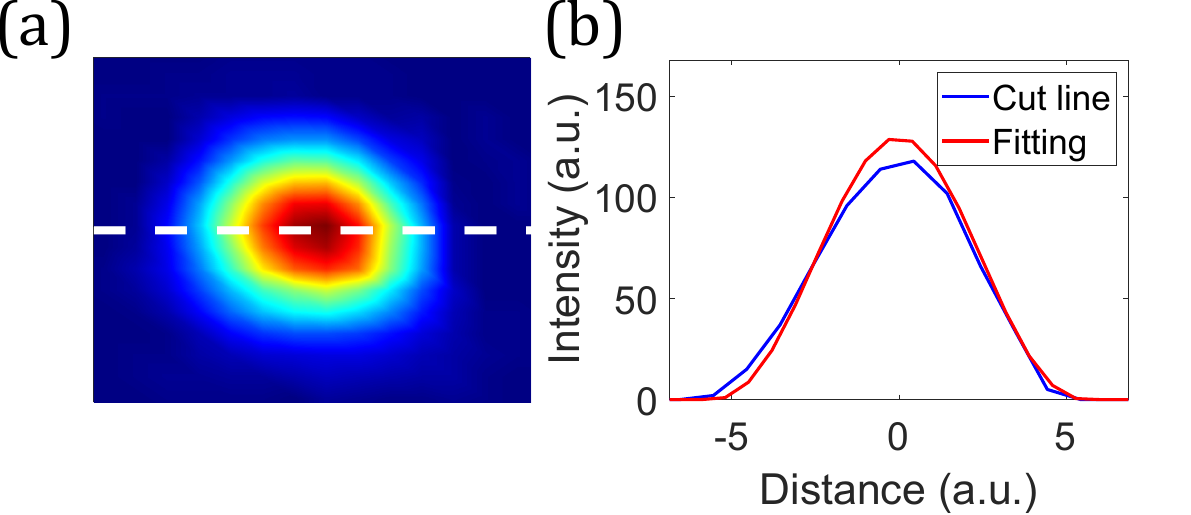}
\caption{The (white) cutline of the spatial profile at \SIadj{40}{\bar} \ce{H2} pressure (a) is shown in (b). The central part is fitted to $c\cdot J_0(k(x-x_0))$, the fundamental transverse mode of a hollow-core fiber.}
\label{fig:output_profile_cutline}
\end{figure}

Fig.~\ref{fig:output_profile} shows the measured output spatial profiles of the filtered Raman beam (Fig.~2) at pressures from \SI{0}{\bar} to \SI{80}{\bar}. In Fig.~\ref{fig:output_profile_cutline}, the spatial profile at \SI{40}{\bar} is fitted to a Bessel function of the first kind as an example to show that these profiles correspond to the fundamental transverse mode of the AR-HCF.


\section{Resonance-induced dispersive-wave generation}

\begin{figure}[h!]
\centering
\includegraphics[width=0.9\linewidth]{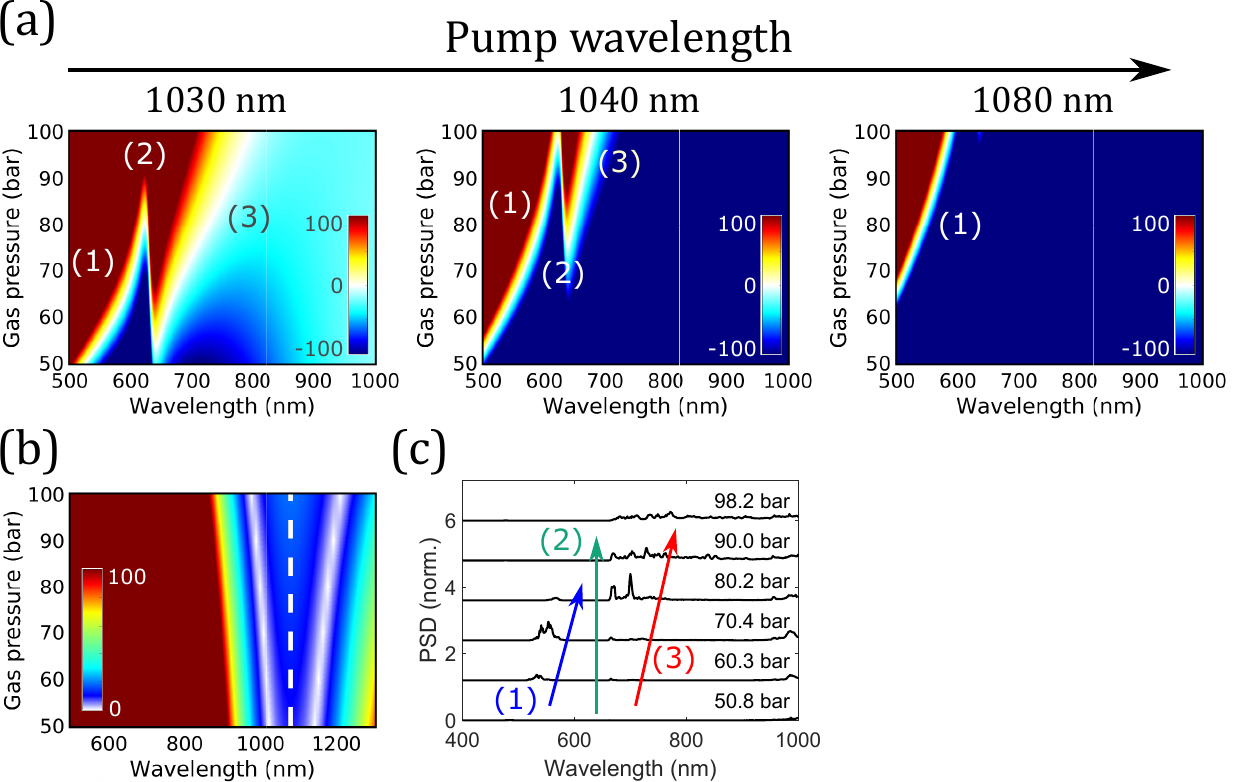}
\caption{(a) The evolution of the phase-matching relation ($\triangle\beta_{\text{DW}}$) of the dispersive wave at different pump wavelengths. (b) The phase-matching relation ($\abs{\triangle\beta_{\text{FWM}}}$) of the four-wave mixing. The white dashed line represents the pump wavelength. (c) The close-view of the visible spectra of Fig.~2(a) in the manuscript. (1-3) are three efficient dispersive-wave emissions where $\triangle\beta_{\text{DW}}\sim0$. The units of both $\triangle\beta$ are \si{1/\m}.}
\label{fig:spectrum_DW}
\end{figure}

Due to the sharp dispersion slope near a resonance, efficient narrowband dispersive-wave generation occurs at high gas pressures despite the high confinement loss \cite{Tani2018,Chen2020a}. Fig.~\ref{fig:spectrum_DW}(a) depicts the phase-matching relation of a dispersive wave, which is calculated from
\begin{equation}
\triangle\beta_{\text{DW}}=\beta(\omega)-\left[\beta(\omega_p)+\beta_1\left(\omega-\omega_p\right)+\beta^{\text{Kerr}}_{\text{DW}}\right],
\end{equation}
where $\beta(\omega)$ is the propagation constant at angular frequency $\omega$ and those in the square brackets represent the propagation constant of the pump pulse. $\beta(\omega_p)$ and $\beta_1$ are the propagation constant and the inverse group velocity at the pump wavelength, $\omega_p$ is the pump angular frequency, $\beta^{\text{Kerr}}_{\text{DW}}=\gamma P_0\omega/\omega_p$ is the nonlinear contribution of the optical Kerr effect, $\gamma=\frac{n_2\omega_p}{cA_{\text{eff}}}$ is the nonlinear coefficient, $P_0$ is the peak power of the pump pulse. Because of the S-shape of the dispersion curve near resonance, there is a maximum of three phase-matched dispersive-wave wavelengths where $\triangle\beta_{\text{DW}}$ approaches zero. At \SI{1030}{\nm}, there are three curves, (1-3) in Fig.~\ref{fig:spectrum_DW}(a), where dispersive waves are phase-matched with the pump pulse. With increasing pump wavelength, only one phase-matching curve is observed. Since our gain-managed pump pulse has a broadband spectrum from \num{1000} to \SI{1100}{\nm}, multiple dispersive waves are generated. In Fig.~\ref{fig:spectrum_DW}(c), we experimentally observed multiple dispersive waves. The first dispersive wave is generated and red-shifts from \num{540} to \SI{570}{\nm} as the gas pressure increases. The second dispersive wave, however, stays around \SI{665}{\nm} with different pressures. Finally, the third one red-shifts as the first one but is rather broadband. These features are consistent with the calculation of the phase-matching relation in Fig.~\ref{fig:spectrum_DW}(a).

Not only dispersive waves but also four-wave mixing (FWM) becomes phase-matched near a resonance. The FWM phase-matching relation is
\begin{equation}
\triangle\beta_{\text{FWM}}=\beta(\omega_s)+\beta(\omega_i)-2\beta(\omega_p)+\beta^{\text{Kerr}}_{\text{FWM}},
\end{equation}
where $\omega_s$ and $\omega_i$ are angular frequencies of the signal and the idler that satisfy $\omega_s+\omega_i-2\omega_p=0$. $\beta^{\text{Kerr}}_{\text{FWM}}=2\gamma P_0$ is the nonlinear contribution of the optical Kerr effect. Fig.~\ref{fig:spectrum_DW}(b) shows the phase-matching relation when the pump wavelength is \SI{1080}{\nm}. The two white curves are the phase-matched signal and the idler. They stay around the pump wavelength and don't contribute to the visible wavelengths. The same feature is found with different pump wavelengths, so they are not shown here.

\section{Discussion of photoionization in scaling of SSFS to higher pulse energy}

To generate a Raman soliton in \ce{H2} through SSFS with high efficiency, a short pulse is preferred. However, to scale to energies of several microjoules, the peak power of the input pulse can be high enough to ionize \ce{H2}. This may lead to phenomena such as photoionization-induced blue-shift \cite{Saleh2011,Saleh2021}. Keldysh theory provides a framework for quantitative analysis of this issue \cite{Keldysh1965}. In particular, the Keldysh parameter $\gamma$ can be used to delineate the transition between the high-intensity tunneling domain ($\gamma\ll1$) and the low-intensity multiphoton ionization ($\gamma\gg1$). For the \SIadj{2}{\micro\joule} pulses considered in the manuscript, the Keldysh parameter is $1.7$, which is similar to the case in \cite{Belli2015}, where ionization does not have a significant effect. $\gamma\approx1$ for \SIadj{5.8}{\micro\joule} pulses in a hydrogen-filled fiber of a \SIadj{30}{\micro\meter} core diameter; in that case, photoionization will be appreciable \cite{Saleh2011,Koettig2017,Saleh2021}. To avoid ionization, Jenkins \etal have proposed use of the divided-pulse technique \cite{Zhou2007} to reduce the peak power  \cite{Jenkins2020}. This technique has also been applied experimentally in SSFS but with a solid-core photonic crystal fiber \cite{Zhang2017a}. Divided-pulse SSFS may be a way to avoid photoionization in scaling SSFS in hydrogen to energies above several microjoules.

\clearpage


\bibliography{reference.bib}

\end{document}